\documentclass[12pt,epsf]{article}

\usepackage{times}
\usepackage{graphicx}
\usepackage{amsfonts}
\usepackage{amsmath}
\usepackage{amssymb}
\usepackage{amstext}
\usepackage{amsthm}

\usepackage{epsfig}
\usepackage{color}
\psfigdriver{dvips}
\usepackage{latexsym}
\usepackage[matrix,curve]{xypic}
\usepackage{rotating}
\rotdriver{dvips}

\begin{document}

\baselineskip 6mm
\renewcommand{\thefootnote}{\fnsymbol{footnote}}

%------------ Hyun Seok's macro's, etc  -----------

\newcommand{\nc}{\newcommand}
\newcommand{\rnc}{\renewcommand}

%\headheight=0truein
%\headsep=0truein
%\topmargin=0truein
%\oddsidemargin=0truein
%\evensidemargin=0truein
%\textheight=9truein
%\textwidth=6.5truein

\rnc{\baselinestretch}{1.24}    % 1.5 spacing btwn text lines
\setlength{\jot}{6pt}       % spacing btwn the rows of an eqnarray
\rnc{\arraystretch}{1.24}   % spacing btwn the rows of a non-eqn array

%%%%%%%%%%%%%%%%%%%%%% Equation Numbering %%%%%%%%%%%%%%%%%%%%%%%
\makeatletter
%\rnc{\theequation}{\thesection.\arabic{equation}}
\@addtoreset{equation}{section}
\makeatother

%%%%%%%%%%%%%%%%%%%%%%%%%%%%%%%%%%%%%%%%%%%%%%%%%%%%%%%%%%%%%%%%%
%                                                               %
%                NEW COMMANDS AND MACROS                        %
%                                                               %
%%%%%%%%%%%%%%%%%%%%%%%%%%%%%%%%%%%%%%%%%%%%%%%%%%%%%%%%%%%%%%%%%

%%%%% Simplify some frequently used LaTeX commands %%%%%

\nc{\be}{\begin{equation}}

\nc{\ee}{\end{equation}}

\nc{\bea}{\begin{eqnarray}}

\nc{\eea}{\end{eqnarray}}

\nc{\xx}{\nonumber\\}

\nc{\ct}{\cite}

\nc{\la}{\label}

\nc{\eq}[1]{(\ref{#1})}

\nc{\newcaption}[1]{\centerline{\parbox{6in}{\caption{#1}}}}

\nc{\fig}[3]{

\begin{figure}
\centerline{\epsfxsize=#1\epsfbox{#2.eps}}
\newcaption{#3. \label{#2}}
\end{figure}
}

%%% Caligraphic letters %%%%

\def\CA{{\cal A}}
\def\CC{{\cal C}}
\def\CD{{\cal D}}
\def\CE{{\cal E}}
\def\CF{{\cal F}}
\def\CG{{\cal G}}
\def\CH{{\cal H}}
\def\CK{{\cal K}}
\def\CL{{\cal L}}
\def\CM{{\cal M}}
\def\CN{{\cal N}}
\def\CO{{\cal O}}
\def\CP{{\cal P}}
\def\CR{{\cal R}}
\def\CS{{\cal S}}
\def\CU{{\cal U}}
\def\CV{{\cal V}}
\def\CW{{\cal W}}
\def\CY{{\cal Y}}
\def\CZ{{\cal Z}}

%%% Double line letters %%%

\def\IB{{\hbox{{\rm I}\kern-.2em\hbox{\rm B}}}}
\def\IC{\,\,{\hbox{{\rm I}\kern-.50em\hbox{\bf C}}}}
\def\ID{{\hbox{{\rm I}\kern-.2em\hbox{\rm D}}}}
\def\IF{{\hbox{{\rm I}\kern-.2em\hbox{\rm F}}}}
\def\IH{{\hbox{{\rm I}\kern-.2em\hbox{\rm H}}}}
\def\IN{{\hbox{{\rm I}\kern-.2em\hbox{\rm N}}}}
\def\IP{{\hbox{{\rm I}\kern-.2em\hbox{\rm P}}}}
\def\IR{{\hbox{{\rm I}\kern-.2em\hbox{\rm R}}}}
\def\IZ{{\hbox{{\rm Z}\kern-.4em\hbox{\rm Z}}}}

%%% Greek letters %%%

\def\a{\alpha}
\def\b{\beta}
\def\d{\delta}
\def\ep{\epsilon}
\def\ga{\gamma}
\def\k{\kappa}
\def\l{\lambda}
\def\s{\sigma}
\def\t{\theta}
\def\w{\omega}
\def\G{\Gamma}

%%%%% Mathematical Symbols

\def\half{\frac{1}{2}}
\def\dint#1#2{\int\limits_{#1}^{#2}}
\def\goto{\rightarrow}
\def\para{\parallel}
\def\brac#1{\langle #1 \rangle}
\def\curl{\nabla\times}
\def\div{\nabla\cdot}
\def\p{\partial}

%%%%% Roman pont in math

\def\Tr{{\rm Tr}\,}
\def\det{{\rm det}}

%%%%% Special Letters

\def\vare{\varepsilon}
\def\zbar{\bar{z}}
\def\wbar{\bar{w}}
\def\what#1{\widehat{#1}}

%%%%% For this paper only

\def\ad{\dot{a}}
\def\bd{\dot{b}}
\def\cd{\dot{c}}
\def\dd{\dot{d}}
\def\so{SO(4)}
\def\bfr{{\bf R}}
\def\bfc{{\bf C}}
\def\bfz{{\bf Z}}

\begin{titlepage}

%---------------- preprint number ---------------

%\hfill\parbox{3.7cm} {{\tt arXiv:1503.00712}}

\vspace{15mm}

\begin{center}
%------------------------ title ------------------------
{\Large \bf Emergent Spacetime: Reality or Illusion?}\footnote{Essay written for the Gravity Research
Foundation 2015 Awards for Essays on Gravitation.}

\vspace{10mm}
%---------------- authors and addresses ----------------

Hyun Seok Yang \footnote{hsyang@kias.re.kr}
\\[10mm]

{\sl School of Physics, Korea Institute for Advanced Study,
Seoul 130-722, Korea}

\end{center}

\thispagestyle{empty}

\vskip2cm

%----------------------- abstract ----------------------

\centerline{\bf ABSTRACT}
\vskip 7mm
\noindent

The contemporary physics has revealed growing evidences that the emergence can be applied
to not only biology and condensed matter systems but also gravity and spacetime.
We observe that noncommutative spacetime necessarily implies emergent spacetime
if spacetime at microscopic scales should be viewed as noncommutative.
Since the emergent spacetime is a new fundamental paradigm for quantum gravity,
it is necessary to reexamine all the rationales to introduce the multiverse hypothesis
from the standpoint of emergent spacetime.
We argue that the emergent spacetime certainly opens a new perspective that may cripple all
the rationales to introduce the multiverse picture. Moreover the emergent spacetime may rescue us
from the doomsday of metastable multiverse as quantum mechanics did
from the catastrophic collapse of classical atoms.
\\

%PACS numbers: 04.60.-m, 11.25.Tq, 04.50.-h

%Keywords: Emergent spacetime, Multiverse, Quantum gravity

\vspace{1cm}

\today

\end{titlepage}

\renewcommand{\thefootnote}{\arabic{footnote}}
\setcounter{footnote}{0}

%\section{Introduction}

History is a mirror to the future. If we do not learn from the mistakes of history,
we are doomed to repeat them.\footnote{George Santayana (1863-1952).}
The physics of the last century had devoted to the study of two pillars: general relativity and
quantum field theory. Although the revolutionary theories of relativity and quantum mechanics
have utterly changed the way we think about Nature and the Universe, new open problems have emerged
which have not yet been resolved within the paradigm of the 20th century physics.
In particular, recent developments in cosmology, particle physics and string theory have led to
a radical proposal that there could be an ensemble of universes that might be completely disconnected
from ours \cite{carr-book}. Of course, it would be perverse to claim that nothing exists beyond the horizon of
our observable universe. However, a painful direction is to use the string landscape or multiverse to explain
some notorious problems in theoretical physics based on the anthropic argument \cite{carr-ellis}.
``And it's pretty unsatisfactory to use the multiverse hypothesis to explain only things
we don't understand."\footnote{Graham Ross in {\it Quanta magazine} ``At multiverse impasse, a new theory
of scale" (August 18, 2014) and {\it Wired.com} ``Radical new theory could kill the multiverse hypothesis."}
Taking history as a mirror, this situation is very reminiscent of the hypothetical luminiferous
ether in the late 19th century. Looking forward to the future,
we may need another turn of the spacetime picture to defend the integrity of physics.

Recent developments in string theory have revealed a remarkable and radical new
picture about gravity. In particular, the AdS/CFT duality
illustrates a typical example of emergent gravity and emergent space because gravity in higher
dimensions is defined by a gravityless field theory in lower dimensions \cite{ads-cft}.
Now we have many examples from string theory in which spacetime is not fundamental but
only emerges as a large distance, classical approximation \cite{hopo-rev}. Therefore,
the rule of the game in quantum gravity is that space and time are an emergent concept.
Since the emergent spacetime is a new fundamental paradigm for quantum gravity and
it is exclusive and irreconcilable with the conventional spacetime picture in general relativity,
it is necessary to reexamine all the rationales to introduce the multiverse hypothesis
from the standpoint of emergent spacetime.
The emergent spacetime will certainly open a new prospect \cite{hsy2015} that may cripple all
the rationales to introduce the multiverse picture.

Since the concept of the multiverse raises deep conceptual issues even to require to change
our view of science itself \cite{carr-book,carr-ellis},
it should be important to ponder on the real status of the multiverse
whether it is simply a mirage developed from an incomplete physics like the ether
in the late 19th century or it is of vital importance even in more complete theories.
We think that the multiverse debate in physics circles has to seriously take the emergent
spacetime into account.

For this purpose, let us assume that {\bf spacetime is an emergent entity from some fundamental objects
in quantum gravity} \cite{q-emg}. This means that we do not assume
the prior existence of spacetime but define a spacetime structure as a solution of an underlying
background-independent theory such as matrix models. We will highlight that
a background-independent theory such as matrix models provides a concrete realization
of the idea of emergent spacetime which has a sufficiently elegant and explanatory power
to defend the integrity of physics against the multiverse hypothesis \cite{hsy2015}.

The multiverse hypothesis has been motivated by an attempt to explain the anthropic fine-tuning
such as the cosmological constant problem \cite{weinberg-87} and boosted by the chaotic and
eternal inflation scenarios \cite{inflation} and the string landscape derived from
the Kaluza-Klein compactification of string theory \cite{slands}.
In summary, we list the main (not exhausting) sources of the multiverse idea \cite{carr-book}:
\begin{enumerate}
\item[A.] Cosmological constant problem.
\item[B.] Chaotic and eternal inflation scenarios.
\item[C.] String landscape.
\end{enumerate}
First of all, we have to point out that these are all based on the traditional spacetime picture.
The cosmological constant problem (A) is the problem in all traditional gravity theories such as
Einstein gravity and modified gravities. So far any such a theory has not succeeded to resolve
the problem A. The inflation scenarios (B) are also based on the traditional gravity theory coupled
to an effective field theory for inflaton(s). Thus, in these scenarios, the prior existence
of spacetime is simply assumed. The string landscape (C) also arises from the conventional
Kaluza-Klein compactification of string theory. The string landscape (C) means that the huge variety
of compactified internal geometries exist, typically, in the range of $10^{500}$ and
almost the same number of four-dimensional worlds with different low-energy phenomenologies
accordingly survive \cite{slands}.

In string theory, there are two exclusive spacetime pictures based on the Kaluza-Klein theory
vs. emergent gravity although they are conceptually in deep discord with each other.
On the one hand, the Kaluza-Klein gravity is defined in higher dimensions as a more superordinate theory
and gauge theories in lower dimensions are derived from the Kaluza-Klein theory via compactification.
Since the Kaluza-Klein theory is just the Einstein gravity in higher dimensions, the prior existence of
spacetime is {\it a priori} assumed. On the other hand, in emergent gravity picture,
gravity in higher dimensions is not a fundamental force but a collective phenomenon
emergent from more fundamental ingredients defined in lower dimensions.
In emergent gravity approach, the existence of spacetime is not {\it a priori} assumed but
the spacetime structure is defined by the theory.
This picture leads to the concept of emergent spacetime.
In some sense, emergent gravity is the inverse of Kaluza-Klein paradigm, schematically summarized by
\begin{equation}\label{kk-em}
    (1 \otimes 1)_S \rightleftarrows 2 \oplus 0
\end{equation}
where $\rightarrow$ means the emergent gravity picture while $\leftarrow$ indicates the Kaluza-Klein picture.

Our leitmotif is that a consistent theory of quantum gravity should be background-independent,
so that it should not presuppose any spacetime background on which fundamental processes develop.
Hence the background-independent theory must provide a mechanism of spacetime generation such that
every spacetime structure including the flat spacetime arises as a solution of the theory itself.
A profound feature in emergent gravity is that even the flat spacetime must
have a dynamical origin, which is absent in general relativity.
It turns out \cite{q-emg} that the flat Minkowski spacetime
is originated from a noncommutative (NC) spacetime which is a vacuum solution in the Coulomb branch of
a large $N$ matrix model. General solutions are generated by considering generic
deformations of the primitive vacuum. A striking fact is that the vacuum responsible for the generation
of flat spacetime is not empty and admits a separable Hilbert space as quantum mechanics.
Amusingly the perverse vacuum energy $\rho_{\mathrm{vac}} \sim  M_P^4$ known as the cosmological constant
in general relativity was actually the origin of flat spacetime.
This is a tangible difference from Einstein gravity, in which $T_{\mu\nu} = 0$ in flat spacetime.
In the end, the emergent gravity reveals a remarkable picture
that the cosmological constant does not gravitate.
To emphasize clearly, the emergent gravity does not contain the coupling of
cosmological constant like $\int d^{4} x \sqrt{-g} \Lambda$, so it presents a surprising
contrast to general relativity. In consequence, the emergent gravity definitely dismisses
the problem A \cite{hsy-jpcs12}. Therefore, there is no demanding reason
to rely on the anthropic fine-tuning to explain the tiny value of current dark energy.

The multiverse picture arises in inflationary cosmology (B) since,
in most inflationary models, once inflation happens, it produces not just one universe,
but an infinite number of universes \cite{inflation}. Thus an important question is whether
the emergent spacetime picture can also lead to the eternal inflation.
The answer is certainly no. The reason is the following \cite{hsy2015}.
The inflationary vacuum in emergent gravity describes the creation of spacetime unlike
the traditional inflationary models that simply describe the exponential expansion of a preexisting spacetime.
Moreover, the inflation corresponds to a dynamical process of the Planck energy condensate
into vacuum responsible for the emergence of spacetime.
An important point is that the Planck energy condensate results in a highly coherent
vacuum called the NC space. As the NC phase space in quantum mechanics
necessarily brings about the Heisenberg's uncertainty relation, $\Delta x \Delta p \geq
\frac{\hbar}{2}$, the NC space also leads to the spacetime uncertainty relation.
Therefore any further accumulation of energy over the vacuum must be subject to
the exclusion principle known as the UV/IR mixing \cite{uv-ir}.
Consequently, it is not possible to further accumulate the Planck energy density $\delta \rho \sim M_P^{4}$
over the inflating spacetime. This means that it is impossible to superpose
a new inflating subregion over the inflationary universe. In other words,
the cosmic inflation triggered by the Planck energy condensate into vacuum must be
a single event \cite{hsy2015}. In the end, we have a beautiful picture: The NC spacetime is necessary 
for the emergence of spacetime and the exclusion principle of NC spacetime guarantees the stability
of spacetime.\footnote{This situation is similar to quantum mechanics rescued us from
the catastrophic collapse of classical atoms. Similarly the emergent spacetime may rescue us from
the doomsday of metastable multiverse as warned in another essay \cite{asen}.}

The above argument suggests an intriguing picture for the dark energy too.
Since the flat spacetime arises from the vacuum obeying the Heisenberg algebra,
any local fluctuations over the NC spacetime must also be subject to the spacetime
uncertainty relation or UV/IR mixing. This implies that any UV fluctuations are paired with
corresponding IR fluctuations. For example, the most typical UV fluctuations are characterized
by the Planck mass $M_P$ and these will be paired with the most typical IR fluctuations with
the largest possible wavelength denoted by $L_H = M_H^{-1}$. This means that these UV/IR fluctuations
are extended up to the scale $L_H$ which may be identified with the size of
our observable universe, $L_H \sim 1.3 \times 10^{26}$ m.
A simple dimensional analysis shows that the energy density of these fluctuations
is roughly equal to the current dark energy, i.e.,
\begin{equation}\label{dark-energy}
\delta \rho \sim  M_{DE}^4 = \frac{1}{L_P^2 L_H^2} \sim (10^{-3} \mathrm{eV})^4.
\end{equation}
Thus the emergent gravity predicts the existence of dark energy whose scale is characterized
by the size of our visible universe \cite{hsy-jpcs12}.

An important point is that the emergence of gravity requires the emergence of spacetime too.
If spacetime is emergent, everything supported on the spacetime should also be emergent.
In particular, matters cannot exist without spacetime and thus must be emergent together with the spacetime.
Eventually, the background-independent theory has to make no distinction between
geometry and matter \cite{q-emg}. This is the reason why the emergent spacetime
cannot coexist peacefully with the Kaluza-Klein picture.
As we pointed out before, the string landscape was derived from the Kaluza-Klein compactification
of string theory. However, if the emergent spacetime picture is correct,
the string landscape may be endowed with a completely new interpretation
since reversing the arrow in \eq{kk-em} accompanies a radical change of physics.
For example, a geometry is now derived from a gauge theory while previously the gauge theory
was derived from the geometry.

The Kaluza-Klein compactification of string theory advocates that the Standard Model in four dimensions
is determined by a six-dimensional internal geometry, e.g., a Calabi-Yau manifold.
Thus different internal geometries mean different physical laws in four dimensions,
so different universes governed by different Standard Models. However,
the emergent gravity reverses the arrow in \eq{kk-em}. Rather internal geometries are determined
by microscopic configurations of gauge fields and matter fields in four dimensions.
As a consequence, different internal geometries mean different microscopic configurations of
four-dimensional particles and nonperturbative objects such as solitons and instantons.
In this picture, the huge variety of internal geometries may correspond to the ensemble of microscopic
configurations in four dimensions and $10^{500}$ would be the Avogadro number for the microscopic ensemble.
Recall that NC geometry begins from the correspondence between the category of topological spaces
and the category of commutative algebras over $\mathbb{C}$ and
then changes the commutative algebras by NC algebras to define corresponding NC spaces.
In this correspondence, different internal geometries correspond to choosing different
NC algebras. A crucial point is that the latter allows a background-independent formulation
which does not require a background geometry.
Hence a background-independent quantum gravity seems to bring a new perspective that
cripples all the rationales to introduce the multiverse hypothesis.

\section*{Acknowledgments}

This work was supported by the National Research Foundation of Korea (NRF) grant
funded by the Korea government (MOE) (No. 2011-0010597).

\end{document}